\documentclass[aps,twocolumn]{revtex4-1}
\usepackage{graphicx}
\usepackage{epstopdf}
\usepackage[caption=false]{subfig}
\begin{document}
\title{General relations between the power, efficiency and dissipation for the irreversible heat engines in the nonlinear response regime}
\author{I. Iyyappan}
\email[]{iyyap.si@gmail.com}
\author{M. Ponmurugan}
\email[]{ponphy@cutn.ac.in}
\affiliation{Department of Physics, School of Basic and Applied Sciences, Central
University of Tamil Nadu, Thiruvarur 610 005, Tamil Nadu, India.}
\begin{abstract}
We derive the general relations between the maximum power, maximum efficiency and minimum dissipation for the irreversible heat engine  in nonlinear response regime. In this context, we use the minimally nonlinear irreversible model and obtain the lower and upper bounds of the above relations for the asymmetric dissipation limits. These relations can be simplified further when the system possesses the time-reversal symmetry or anti-symmetry. We find that our results are the generalization of various such relations obtained earlier for different heat engines.
\end{abstract}
\maketitle
\section{Introduction}
The second law of thermodynamics limits the complete conversion of heat into work while allowing the complete conversion of work into heat \cite{call}. For practical applications such a conversion from one of its form into another is very important, which can be done by the heat devices. The heat engines absorb the input heat $Q_{h}$ from the hot reservoir at temperature $T_h$, converts part of it into the useful work $W$ and eject the remaining amount of heat $Q_{c}$ to the cold reservoir at temperature $T_c$ ($T_h>T_c$). The heat engine efficiency defined as, $\eta=W/Q_{h}$, reaches the Carnot efficiency, $\eta_C=1-T_c/T_h$ when it working reversibly \cite{carn}. Since the reversible process takes an infinite amount of time to attain the Carnot efficiency, the power output becomes zero and hence does not have any practical use. The $\eta_C$ is the universal upper bound for the efficiency of the heat engines working between the two reservoirs at different temperatures.
	
The efficiency at maximum power is an another important criterion for heat engines performance, which were widely studied in the literature for the macroscopic heat engines \cite{curz22,chen114,vela100,espo130,van190,tu312,espo150,espo041,wang011,aper041,shen402,ayal052,ryab050}, stochastic heat engines \cite{schm200,holu050,rana042,verl472,ging102,holu052,proe065,proe023,raz160,lee161,proe041,proe170}, the quantum heat engines \cite{lin046,espo600,nakp235,wang031,harb500,gosw013,camp035} and the heat engines with the finite-size reservoirs \cite{izum180,wang062,wang012,ponm160,joha100,joha012}. The efficiency at maximum power also has the universal nature up to the second order in Carnot efficiency \cite{espo130,tu312,schm200} as $\eta_{MP}=\eta_C/2+\eta_C^{2}/8+O(\eta_C^{3})$.

Finding the universal upper bound for the efficiency at maximum power is one of the main goal in finite-time thermodynamics. The total  entropy production $S$, needs to be minimized to increase the efficiency. The dissipation of the heat engines can be written as a function of the efficiency $\eta$ and power output $P$ as \cite{stru}
\begin{equation}\label{2b}
T_c\dot{S}=P\left(\frac{\eta_C}{\eta}-1\right).
\end{equation}
When the entropy production rate $\dot{S}$ becomes zero, the heat engines reach the Carnot efficiency. In recent years, many authors were attempted to investigate the attainability of a Carnot efficiency with finite power output \cite{bene230,alla050,whit130,pole050,proe090,bran031,john170}, in particular, the heat engines with broken time-reversal symmetry were shown to enhance the efficiency at maximum power \cite{bene230,sait201,sanc201,bran070,bala165,bran105,bran012,yama121}. Also the general relations between the maximum power ($P_{MP}$), the efficiency at maximum power ($\eta_{MP}$), the maximum efficiency ($\eta_{ME}$) and power at maximum efficiency ($P_{ME}$) were identified for the specific models of heat engines \cite{jian042,baue042}. 

Recently, Proesmans \textit{et al}. obtained such a general relations for the linear irreversible heat engine with and without  time-reversal symmetry (anti-symmetry) \cite{proe220}. These relations showed that the efficiency at maximum power is bounded by the half the Carnot efficiency for the system possessing time-reversal symmetry or anti-symmetry. However, several other studies showed that the presence of nonlinearity in the system can increase such bound \cite{izum100,long062}. In such a case there is a need for finding out the generalized relations valid also in the nonlinear regime. This urge us to study the general relation for the irreversible heat engine in the nonlinear regime. In this context, we use the minimally nonlinear model in the generalized framework which directly include the heat dissipation in the irreversible heat engine in the nonlinear regime \cite{izum100}. For the asymmetric dissipation limits with the tight-coupling condition (i.e., is no heat leakage between the system and reservoir), in this model, one can obtain the lower and upper bounds of the efficiency at maximum power as \cite{izum100}
\begin{equation}
\frac{\eta_C}{2}\leq\eta_{MP}\leq\frac{\eta_C}{2-\eta_C}.
\end{equation}
These bounds were first derived for the low-dissipation Carnot heat engines with the asymmetrical dissipation limits \cite{espo150}. The minimally nonlinear irreversible heat engine model was widely studied in the literature \cite{wang012,ponm160,long062,shen012,izum052,long052} and shown that it is equivalent to the low-dissipation Carnot heat engine, latter explains the performance of the real power plants very well  \cite{izum100,long062,long052,de012,holu073}. 

In this paper, we derive the general relations between the maximum power, maximum efficiency and minimum dissipation for the minimally nonlinear irreversible heat engines in the generalized framework. We find that the results obtained for the linear irreversible heat engines are the special case of our general results obtained for the asymmetrical dissipation limit. Interestingly, we also find that our results are the generalization of various such relations obtained earlier for the different heat engines.

This paper is organized as follows, in section II, we briefly review the general relations for the linear irreversible heat engine. In section III, we study the minimally nonlinear irreversible heat engine and derive our main results. In section IV, we discuss about our results. We end with our conclusion in section V.     
\section{The linear irreversible heat engine}
If the system performs the work on the environment, $W = -F x$, where $F$ is the constant external force and $x$ is the thermodynamic variable conjugate of $F$. The power becomes, $P = -F\dot{x}$, where the dot denotes the time derivative of the quantity. The input and output heat fluxes are, respectively, $\dot{Q}_h$ and $\dot{Q}_c$. The entropy production rate at the reservoirs is given by \cite{van190,izum052}
\begin{equation}\label{2c}
\dot{S}=-\frac{\dot{Q}_h}{T_h}+\frac{\dot{Q}_c}{T_c}=-\frac{F\dot{x}}{T_c}+\dot{Q}_h\left(\frac{1}{T_c}-\frac{1}{T_h}\right).
\end{equation}   
We can write the above entropy production rate in terms of the thermodynamic forces $\textbf{X}$ and fluxes $\textbf{J}$ as, $\dot{S}= J_1X_1+J_2X_2$ \cite{onsa226,call}. Where $X_1\equiv -F/T_c$ is the thermodynamic force for work and its corresponding thermodynamic flux $J_1\equiv\dot{x}$ and $X_2 \equiv 1/T_c - 1/T_h$ is the thermodynamic force for heat flow and its corresponding thermodynamic flux $J_2\equiv\dot{Q}_h$, which is the input heat flux absorbed by the system from the hot reservoir at temperature $T_h$. When the thermodynamic forces are small, the thermodynamic fluxes can be written in the linear combination of the thermodynamic forces as \cite{onsa226,call}
\begin{equation}\label{2d}
J_1=L_{11}X_1+L_{12}X_2,
\end{equation}
\begin{equation}\label{2e}
J_2=L_{21}X_1+L_{22}X_2,
\end{equation}
where $L_{ij}(i,j=1,2)$ are the Onsager coefficients. The power output can be written as $P=-T_cJ_1X_1$ and the efficiency is defined as $\eta=P/\dot{Q}_h$ and it can be written as \cite{van190} 
\begin{equation}
\eta=\frac{-T_c(L_{11}X_1^{2}+L_{12}X_1X_2)}{L_{21}X_1+L_{22}X_2}.
\end{equation}
Remarkably, the efficiency at maximum power, the minimum dissipation ($\dot{S}_{mD}$) and the power at minimum dissipation ($P_{mD}$) is linked by the following simple relations \cite{jian042,baue042,proe220}
\begin{equation}\label{2n}
\eta_{MP}=\frac{P_{MP}}{2P_{MP}-P_{ME}}\eta_{ME},
\end{equation}
\begin{equation}\label{2o}
T_c\dot{S}_{mD}=\left(\frac{\eta_C}{\eta_{MP}}-\frac{\eta_C^{2}}{\eta_{ME}^{2}}-1\right)P_{MP}+\frac{\eta_C}{\eta_{ME}^{2}}P_{ME},
\end{equation}
\begin{equation}\label{2p}
P_{mD}=P_{MP}-\frac{\eta_C}{\eta_{ME}^{2}}(P_{MP}-P_{ME}),
\end{equation}
where the Carnot efficiency $\eta_C=T_cX_2$. When the Onsager coefficients satisfy the symmetric or anti-symmetric property ($L_{12}=\pm L_{21}$), one can obtain the further simple relations as \cite{jian042,baue042,proe220}
\begin{equation}\label{2q}
\frac{P_{ME}}{P_{MP}}=\frac{\eta_C^{2}-\eta_{ME}^{2}}{\eta_C^{2}},
\end{equation}
\begin{equation}\label{2r}
\eta_{MP}=\frac{\eta_C^{2}\eta_{ME}}{\eta_C^{2}+\eta_{ME}^{2}},
\end{equation}
\begin{equation}\label{2da}
P_{mD}=0, \quad T_c\dot{S}_{mD}=P_{MP}\left(\frac{1}{\eta_{MP}}-\frac{2}{\eta_C}\right).
\end{equation}
From Eq. (\ref{2q}), we can find that the maximum efficiency $\eta_{ME}$ attains the Carnot efficiency only when the $P_{ME}$ is zero. Therefore, the Carnot efficiency is unattainable with the finite power for the linear irreversible heat engine when the Onsager coefficients possess the time-reversal symmetric (anti-symmetric) property  \cite{proe220} and from Eq. (\ref{2r}) the efficiency at maximum power, $\eta_{MP}\rightarrow\eta_C/2$ when $\eta_{ME}\rightarrow \eta_C$. In addition, we can identify from Eq. (\ref{2da}) is that when the minimum dissipation $\dot{S}_{mD}=0$, the efficiency at maximum power bounded by half the Carnot efficiency for the maximum power $P_{MP}>0$ \cite{proe220}.
\section{The minimally nonlinear irreversible heat engine}
The heat dissipation is inevitable in the real heat engines. In order to include the heat dissipation in linear irreversible thermodynamics, Izumida and Okuda were added a second-order nonlinear term in the linear Onsager relations and assuming the other higher order terms are negligible \cite{izum100}, which thus called as the minimally nonlinear irreversible heat engine. The extended Onsager relations for the minimally nonlinear irreversible heat engine is given by \cite{izum100}
\begin{equation}\label{2s}
J_1=L_{11}X_1+L_{12}X_2,
\end{equation}
\begin{equation}\label{2t}
J_2=L_{21}X_1+L_{22}X_2-\gamma_h J_1^{2},
\end{equation}
where $\gamma_h>0$ is the strength of the dissipation between the hot reservoir and system. In what follows, we will analyze the heat engine in the general setting without assuming any symmetry in the Onsager coefficients ($L_{12}\neq\pm L_{21}$). The power output  can be written as \cite{izum100}
\begin{equation}\label{2u}
P=\frac{L_{12}}{L_{11}}\eta_C J_1-\frac{T_c}{L_{11}}J_1^{2}.
\end{equation}
It has to be noted that the power of the minimally nonlinear irreversible heat engines do not depend on the heat dissipation introduced in the input heat flux \cite{izum100}. Using Eq. (\ref{2s}) we can rewrite the input heat flux as \cite{izum100,izum052} 
\begin{equation}\label{2v}
J_2=\frac{L_{21}}{L_{11}} J_1+\frac{\mathcal{D}}{L_{11}}X_2-\gamma_h J_1^{2},
\end{equation}
where $\mathcal{D}\equiv L_{11}L_{22}-L_{12}L_{21}$. The heat flux ejected to the cold reservoir $\dot{Q}_c\equiv J_3=J_2-P$. Using Eqs. (\ref{2u}) and (\ref{2v}), we get
\begin{equation}\label{2w}
J_3=\frac{L_{21}-L_{12}\eta_C}{L_{11}} J_1+\frac{\mathcal{D}}{L_{11}}+\gamma_c J_1^{2},
\end{equation}
where $\gamma_c\equiv T_c/L_{11}-\gamma_h>0$ is the strength of the dissipation between the cold reservoir and system. The efficiency can be written as \cite{izum100,izum052}
\begin{equation}\label{2y}
\eta=\frac{L_{12}\eta_C J_1-T_c J_1^{2}}{L_{21} J_1+\mathcal{D}X_2-\gamma_h L_{11} J_1^{2}}.
\end{equation}
Maximizing the power (Eq. (\ref{2u})) with respect to $J_1$, we get the maximum power at $J_1^{MP}=L_{12}\eta_C/(2T_c)$ as  \cite{izum100}
\begin{equation}\label{2z}
P_{MP}=\frac{\eta_C L_{12}^{2}}{4L_{11}}X_2.
\end{equation}
The efficiency at maximum power becomes \cite{izum100}
\begin{equation}\label{2aa}
\eta_{MP}=\frac{\eta_C L_{12}^{2}}{4\mathcal{D}+2L_{12}L_{21}-\gamma L_{12}^{2}\eta_C},
\end{equation}
where the dimensionless parameter $\gamma\equiv 1/(1+\gamma_c/\gamma_h)$ provides the dissipation strength in terms of the power dissipations ratio ($\gamma_c/\gamma_h$) between the cold and hot reservoirs. For the asymmetric dissipation limits $\gamma_c/\gamma_h\rightarrow \infty$ and $\gamma_c/\gamma_h\rightarrow 0$, we get the corresponding values of $\gamma$ as, $\gamma\rightarrow 0$ and $\gamma\rightarrow 1$, respectively. For the symmetrical dissipation case $\gamma_h=\gamma_c$, we get $\gamma=1/2$.  Maximizing the efficiency (Eq. (\ref{2y})) with respect to $J_1$, we find the efficiency achieves its maximum at
\begin{equation}\label{2ab}
J_1^{ME}=\frac{-\mathcal{D}+f_D}{L_{21}-\gamma L_{12}\eta_C}X_2,
\end{equation}
where $f_D=\sqrt{\mathcal{D}(L_{11}L_{22}- \gamma L_{12}^{2}\eta_C)}$. Substituting Eq. (\ref{2ab}) in Eqs. (\ref{2u}) and (\ref{2y}), we get the power at maximum efficiency and the maximum efficiency, respectively, as
\begin{equation}\label{2ac}
P_{ME}=-\frac{\eta_C (\mathcal{D}-f_D)\left[
L_{11}L_{22}-\gamma L_{12}^{2}\eta_C-f_D\right]}{L_{11}(L_{21}-\gamma L_{12}\eta_C)^{2}}X_2,
\end{equation}
\begin{equation}\label{2ad}
\eta_{ME}=\frac{\eta_C L_{12}^{2}}{2\mathcal{D}+L_{12}L_{21}+2f_D}.
\end{equation}
Substituting Eqs. (\ref{2u}) and (\ref{2v}) in Eq. (\ref{2b}), the dissipation becomes
\begin{equation}\label{2ae}
\dot{S}=\frac{(L_{21}-L_{12}) X_2 J_1+\mathcal{D} X_2^{2}+(1-\gamma\eta_C)J_1^{2}}{L_{11}}.
\end{equation}
Minimizing the above dissipation with respect to $J_1$, we get the minimum dissipation at $J_1^{mD}=(L_{12}-L_{21})\eta_C/[2T_c(1-\gamma\eta_C)]$ as 
\begin{equation}\label{2af}
\dot{S}_{mD}=\frac{[L_{12}^{2}+L_{21}^{2}+4\gamma\eta_C\mathcal{D}-4\mathcal{D}-2L_{12}L_{21}]}{4L_{11}(1-\gamma\eta_C)}X_2^{2}.
\end{equation}
Substituting the $J_1^{mD}$ in Eq. (\ref{2u}), we get the power at minimum dissipation as
\begin{equation}\label{2ag}
P_{mD}=\frac{(L_{12}-L_{21})[L_{12}(1-2\gamma\eta_C)+L_{21}]}{4L_{11}(1-\gamma\eta_C)^{2}}.
\end{equation}
It is straight forward to obtain the $L_{11}$, $L_{22}$ and $L_{12}$, respectively, by using Eq. (\ref{2ad}), Eq. (\ref{2ac}) and Eq. (\ref{2z}). Substituting the values of the  Onsager coefficients $L_{11}$, $L_{22}$ and $L_{12}$ in Eqs. (\ref{2aa}), (\ref{2af}) and (\ref{2ag}), we get the following simple general relations as 
\begin{equation}\label{2ah}
\eta_{MP}=\frac{P_{MP}}{(2-\gamma \eta_{ME})P_{MP}-(1-\gamma \eta_{ME})P_{ME}}\eta_{ME},
\end{equation}
\begin{widetext}
\begin{equation}\label{2al}
T_c\dot{S}_{mD}=\left(\frac{\eta_C}{(1-\gamma\eta_C)\eta_{MP}}-\frac{\eta_C^{2}}{(1-\gamma\eta_C)\eta_{ME}^{2}}-1\right)P_{MP}+\frac{(1-\gamma \eta_{ME})}{(1-\gamma\eta_C)}\frac{\eta_C}{\eta_{ME}^{2}}P_{ME},
\end{equation}
\begin{equation}\label{2an}
P_{mD}=\frac{\eta_C^{2}(1-\gamma\eta_{ME})^{2}P_{ME}+[\eta_C(2\gamma\eta_{ME}-1)-\eta_{ME}](\eta_C-\eta_{ME})P_{MP}}{(1-\gamma\eta_C)^{2}\eta_{ME}^{2}}.
\end{equation}
\end{widetext}
This is our first main result. The above relations are the generalization of Eqs. (\ref{2n}), (\ref{2o}) and (\ref{2p}) which were obtained for the linear irreversible heat engine \cite{proe220}. Further, for the time-reversal symmetric or anti-symmetric cases $L_{12}=\pm L_{21}$, we get the following relations as a generalization of Eqs. (\ref{2q}) and (\ref{2r}) is given by
\begin{equation}\label{2as}
\frac{P_{ME}}{P_{MP}}=\frac{(\eta_C-\eta_{ME})[\eta_{ME}+\eta_C(1-2\gamma\eta_{ME})]}{\eta_C^{2}(1-\gamma\eta_{ME})^{2}},
\end{equation}
\begin{equation}\label{2aq}
\eta_{MP}=\frac{\eta_C^{2}(1-\gamma \eta_{ME})\eta_{ME}}{\eta_C^{2}(1-\gamma\eta_{ME})+(1-\gamma\eta_C)^{2}\eta_{ME}^{2}},
\end{equation}
\begin{equation}\label{2df}
P_{mD}=0,\quad T_c\dot{S}_{mD}=\left(\frac{1}{\eta_{MP}}-\frac{2}{\eta_C}+\gamma\right)P_{MP}.
\end{equation}
This is our second main result. Since the relations (Eqs. (\ref{2ah})-(\ref{2df})) are depends on the ratio between the strength of the dissipations, we will now discuss the extreme cases of the asymmetric, symmetric and minimum dissipation regimes.
\subsection{Asymmetric dissipation ($\gamma_h\neq\gamma_c$)}
When the strength of the dissipation between the system and hot reservoir $\gamma_h\rightarrow 0$, then $\gamma\rightarrow 0$, we get the following lower bound (denoted by superscript $-$) as 
\begin{equation}\label{2ai}
\eta_{MP}^{-}=\frac{P_{MP}}{2P_{MP}-P_{ME}}\eta_{ME},
\end{equation}
\begin{equation}\label{2am}
(T_c\dot{S}_{mD})^{-}=\left(\frac{\eta_C}{\eta_{MP}}-\frac{\eta_C^{2}}{\eta_{ME}^{2}}-1\right)P_{MP}+\frac{\eta_C}{\eta_{ME}^{2}}P_{ME},
\end{equation}
\begin{equation}\label{2ao}
 P_{mD}^{-}=P_{MP}-\frac{\eta_C^{2}}{\eta_{ME}^{2}}(P_{MP}-P_{ME}).
\end{equation}
The above relations are same as the general relation obtained in the linear irreversible heat engine (see Eqs. (\ref{2n}) - (\ref{2p})). Further, for the time-reversal symmetric or anti-symmetric case, we get the following relations as in Eqs. (\ref{2q}) and (\ref{2r}) as \cite{jian042,baue042,proe220}
\begin{equation}\label{2at}
\left(\frac{P_{ME}}{P_{MP}}\right) ^{-}=\frac{\eta_C^{2}-\eta_{ME}^{2}}{\eta_C^{2}},
\end{equation}
\begin{equation}\label{2ar}
\eta_{MP}^{-}=\frac{\eta_C^{2}\eta_{ME}}{\eta_C^{2}+\eta_{ME}^{2}},
\end{equation}
\begin{equation}\label{2cv}
\left(T_c\dot{S}_{mD}\right)^{-}= P_{MP}\left(\frac{1}{\eta_{MP}}-\frac{2}{\eta_C}\right).
\end{equation}
When the strength of the dissipation between the system and cold reservoir $\gamma_c\rightarrow 0$, then $\gamma\rightarrow 1$, we get the following upper bound (denoted by superscript $+$) as 
\begin{equation}\label{2ai2}
\eta_{MP}^{+}=\frac{P_{MP}}{(2- \eta_{ME})P_{MP}-(1-\eta_{ME})P_{ME}}\eta_{ME},
\end{equation}
\begin{widetext}
\begin{equation}\label{2am2}
(T_c\dot{S}_{mD})^{+}=\frac{1}{(1-\eta_C)}\left(\frac{\eta_C}{\eta_{MP}}-\frac{\eta_C^{2}}{\eta_{ME}^{2}}-1+\eta_C\right) P_{MP}+\frac{(1- \eta_{ME})}{(1-\eta_C)}\frac{\eta_C}{\eta_{ME}^{2}}P_{ME},
\end{equation}
\begin{equation}\label{2ao2}
P_{mD}^{+}=\frac{\eta_C^{2}(\eta_{ME}-1)^{2}P_{ME}+[\eta_C(2\eta_{ME}-1)-\eta_{ME}](\eta_C-\eta_{ME})P_{MP}}{(1-\eta_C)^{2}\eta_{ME}^{2}}.
\end{equation}
\end{widetext}
For the time-reversal symmetric or anti-symmetric case, we get the following relations
\begin{equation}\label{2at2}
\left(\frac{P_{ME}}{ P_{MP}}\right)^{+}=\frac{(\eta_C-\eta_{ME})[\eta_{ME}+\eta_C(1-2\eta_{ME})]}{\eta_C^{2}(1-\eta_{ME})^{2}},
\end{equation}
\begin{equation}\label{2ar2}
\eta_{MP}^{+}=\frac{\eta_C^{2}(1-\eta_{ME})\eta_{ME}}{\eta_C^{2}(1-\eta_{ME})+(1-\eta_C)^{2}\eta_{ME}^{2}},
\end{equation}
\begin{equation}\label{2ch}
\left(T_c\dot{S}_{mD}\right)^{+}= P_{MP}\left(\frac{1}{\eta_{MP}}-\frac{2}{\eta_C}+1\right).
\end{equation}
From Eqs. (\ref{2at}) and (\ref{2at2}), we find that the maximum efficiency $\eta_{ME}$ attains the Carnot efficiency only when $P_{ME}=0$, which supports the previous result obtained by Proesmans \textit{et al}. for the linear irreversible heat engine \cite{proe220}. Also from Eqs. (\ref{2ar}) and (\ref{2ar2}), we get the lower and upper bounds of the  efficiency at maximum power as $\eta_C/2<\eta_{MP}<\eta_C/(2-\eta_C)$, which was obtained for low-dissipation heat engine and minimally nonlinear irreversible heat engine with the tight-coupling condition in the asymmetric dissipation limits \cite{espo150,izum100}.
\subsection{Symmetric dissipation ($\gamma_h= \gamma_c$)}
For the symmetric dissipation case where $\gamma_h=\gamma_c$, then $\gamma= 1/2$, we get the following relations (denoted by superscript $sym$) as
\begin{equation}\label{2ca}
\eta_{MP}^{sym}=\frac{2P_{MP}}{(4- \eta_{ME})P_{MP}-(2-\eta_{ME})P_{ME}}\eta_{ME},
\end{equation}
\begin{widetext}
\begin{equation}\label{2cb}
(T_c\dot{S}_{mD})^{sym}=\left(\frac{2\eta_C}{(2-\eta_C)\eta_{MP}}-\frac{2\eta_C^{2}}{(2-\eta_C)\eta_{ME}^{2}}-1\right)P_{MP}+\frac{(2- \eta_{ME})}{(2-\eta_C)}\frac{\eta_C}{\eta_{ME}^{2}}P_{ME},
\end{equation}
\begin{equation}\label{2cd}
P_{mD}^{sym}=\frac{\eta_C^{2}(2-\eta_{ME})^{2}P_{ME}+4[\eta_C(\eta_{ME}-1)-\eta_{ME}](\eta_C-\eta_{ME})P_{MP}}{(2-\eta_C)^{2}\eta_{ME}^{2}}.
\end{equation}
\end{widetext}
Further, for the time-reversal symmetric or anti-symmetric case, we get the following relations
\begin{equation}\label{2as2}
\left(\frac{P_{ME}}{ P_{MP}}\right)^{sym}=\frac{4(\eta_C-\eta_{ME})[\eta_{ME}+\eta_C(1-\eta_{ME})]}{\eta_C^{2}(2-\eta_{ME})^{2}},
\end{equation}
\begin{equation}\label{2aq2}
\eta_{MP}^{sym}=\frac{2\eta_C^{2}(2- \eta_{ME})\eta_{ME}}{2\eta_C^{2}(2-\eta_{ME})+(2-\eta_C)^{2}\eta_{ME}^{2}},
\end{equation}
\begin{equation}
\left(T_c\dot{S}_{mD}\right)^{sym}= P_{MP}\left(\frac{1}{\eta_{MP}}-\frac{2}{\eta_C}+\frac{1}{2}\right).
\end{equation}
The efficiency at maximum power obtained for an endoreversible heat engine is given by the so called Curzon-Ahlborn (CA) efficiency  $\eta_{CA}=1-\sqrt{T_c/T_h}=1-\sqrt{1-\eta_C}$ \cite{curz22} and it can be expanded for a small temperature difference as $\eta_{CA}\approx\eta_C/2+\eta_C^{2}/8+\eta_C^{3}/16+...$ \cite{izum052}. From Eq. (\ref{2aq2}), we get $\eta_{MP}^{sym}=\eta_C/(2-\eta_C/2)$ when the $\eta_{ME}=\eta_C$. For a small temperature difference between the reservoirs, the $\eta_{MP}^{sym}$ can be expanded as, $\eta_{MP}^{sym}\approx\eta_C/2+\eta_C^{2}/8+\eta_C^{3}/32+...$. This shows that for the symmetric dissipation limit our result equals the $\eta_{CA}$ up to the second order in Carnot efficiency  \cite{izum052}, indicating the universality of the efficiency at maximum power.
\subsection{Minimum dissipation ($\dot{S}_{mD}=0$)}
For heat engines, loss of heat due to the friction between the piston and cylinder and heat transfer between the reservoir and heat engine are impossible to eliminate completely. Hence, the heat dissipation $\dot{S}_{mD}>0$ always for real heat engines. Since the dissipation decreases the efficiency, it needs to be minimized as much as possible in the design and operation of real heat engines.

When the heat engine working at the minimum dissipation $\dot{S}_{mD}=0$ with the maximum efficiency $\eta_{ME}=\eta_C$ and the power at maximum efficiency $P_{ME}=0$, we get from Eq. (\ref{2al}), the efficiency at maximum power ($P_{MP}>0$) as 
\begin{equation}\label{2ak}
\eta_{MP}^{mD}=\frac{\eta_C}{2-\gamma\eta_C}.
\end{equation}
This result was previously obtained for the efficiency at maximum power of a stochastic heat engine by Schmiedl and Seifert at the asymmetric dissipation limits \cite{schm200}, for a thermoelectric heat engine by Apertet \textit{et al}. \cite{aper041} and the minimally nonlinear irreversible heat engine with tight-coupling condition by Izumida \textit{et al}. \cite{izum052}. For the asymmetric dissipation limits $\gamma\rightarrow 0$ and $\gamma\rightarrow 1$, we get the lower and upper bounds as  
\begin{equation}\label{2aj}
\frac{\eta_C}{2}\leq\eta_{MP}^{mD}\leq\frac{\eta_C}{2-\eta_C}.
\end{equation} 
As said above, these bounds were also obtained for the low-dissipation Carnot heat engine and the minimally nonlinear irreversible heat engine with tight-coupling at the asymmetrical dissipation limits \cite{espo150,izum100}. It has to be noted that, in the minimum dissipation limit $\dot{S}_{mD}=0$, we get the above bounds without invoking any symmetry on the Onsager coefficients. We can also get the above bounds from Eq. (\ref{2ah}) for the limiting case $\eta_{ME}=\eta_C$ and $P_{ME}=0$, which shows the interconnection between the Eq. (\ref{2ah}) and Eq. (\ref{2al}) as first observed for linear irreversible heat engine in Ref. \cite{proe220}. For the symmetric dissipation limit $\gamma=1/2$, we get $\eta_{MP}^{mD}=\eta_C/(2-\eta_C/2)$ as obtained earlier in the case of stochastic heat engine for the symmetric dissipation case \cite{schm200}.

\section{Results and Discussion}

\onecolumngrid
\begin{widetext}
\begin{figure}
	\centering
	\subfloat[]{\includegraphics[width=6.0cm]{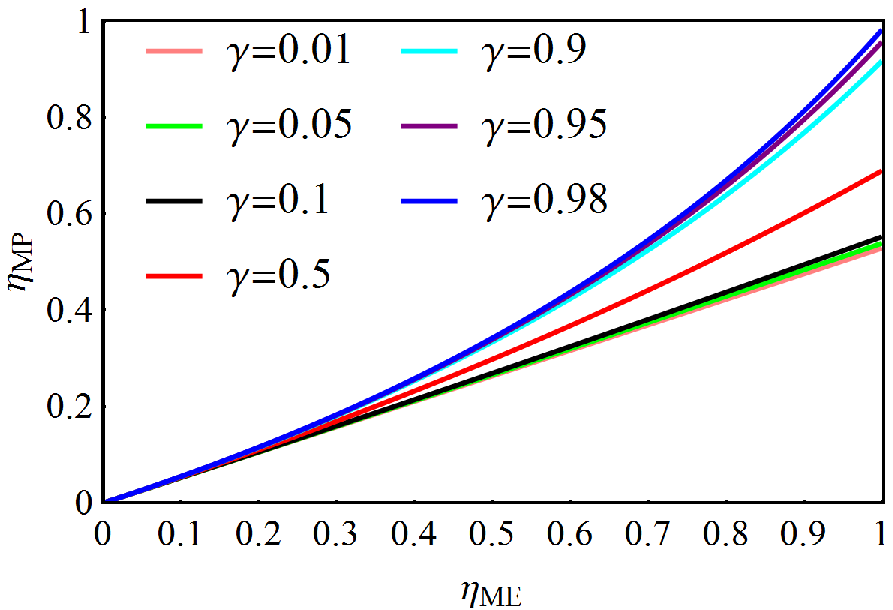}}
	\subfloat[]{\includegraphics[width=6.0cm]{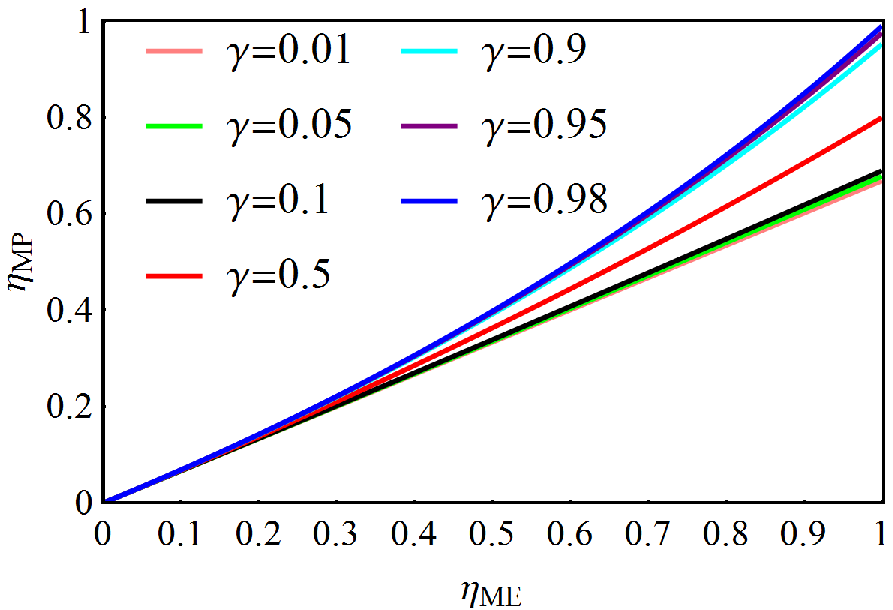}}
	\subfloat[]{\includegraphics[width=6.0cm]{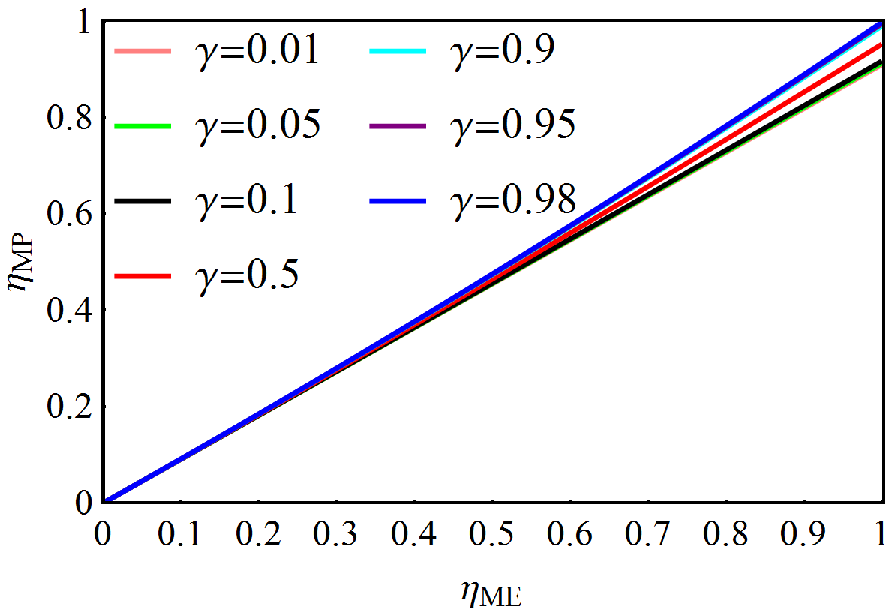}}
	\caption{The efficiency at maximum power Eq. (\ref{2ah}) is plotted as function of the maximum efficiency for various values of $\gamma$ and for different values of power ratio ($P_{ME}/P_{MP}$). In the above figures, the 
values of $P_{ME}/P_{MP}$ taken as in (a) $0.01$, (b) $0.05$ and (c) $0.9$. Here, the pink, green and black curves, respectively,  for $\gamma=0.01$, $\gamma=0.05$, $\gamma=0.1$. Red curve for symmetrical dissipation $\gamma=0.5$. The cyan, purple and blue curves, respectively, for $\gamma=0.9$, $\gamma=0.95$ and  $\gamma=0.98$.}
	\label{fig:etame}
\end{figure}
\end{widetext}
\twocolumngrid

The $\eta_{MP}$ (Eq. (\ref{2ah})) is plotted as function of the $\eta_{ME}$ for different values of $\gamma$ and power ratios $P_{ME}/P_{MP}$ as shown in Fig. \ref{fig:etame}. From Figs. \ref{fig:etame} (a) and (b),  we observe that at lower power ratios, the $\eta_{MP}$ increases linearly with $\eta_{ME}$ for lower values of $\gamma$. Additionally, the linear behavior of $\eta_{MP}$ with $\eta_{ME}$ is also observed for any values of $\gamma$ for the lowest values of $\eta_{ME}$. However, Fig. \ref{fig:etame} (c) shows that the $\eta_{MP}$ varies linearly with $\eta_{ME}$ for all values of $\gamma$ for the higher power ratios.

\begin{figure}[hbtp]
	\includegraphics[width=8.7cm]{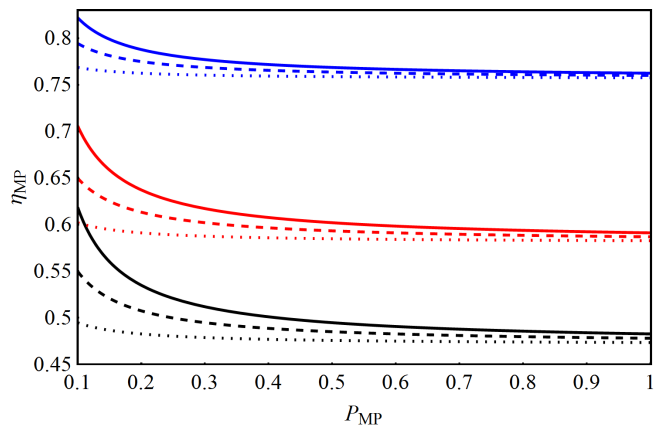}
	\caption{The efficiency at maximum power Eq. (\ref{2ah}) is plotted as function of the maximum power for various values of $\gamma=0.1$ (black), $\gamma =0.5$ (red) and $\gamma=0.9$ (blue) with $P_{ME}=0.01$ (dotted), $0.02$ (dashed) and $0.03$ for (solid). Here, we set $\eta_{ME}=0.9$.}
	\label{fig:pmp}
\end{figure}

To analyze the effect of $P_{MP}$ on $\eta_{MP}$, we plotted $\eta_{MP}$ as function of $P_{MP}$, for different values of $\gamma$ and $P_{ME}$ for a fixed $\eta_{ME}$ in Fig. \ref{fig:pmp}. Even though the $\eta_{MP}$ increases significantly for small increase in $P_{ME}$ for a given $\gamma$, the efficiency at maximum power decreases as the maximum power increases. 

Our results shows that whenever the power ratio approaches unity the $\eta_{MP}$ attains its maximum, which can be increased further when $\gamma$ increases. This can be useful for designing a heat engine to get the higher $\eta_{MP}$ for practical applications.

\begin{figure}[hb]
	\includegraphics[width=8.7cm]{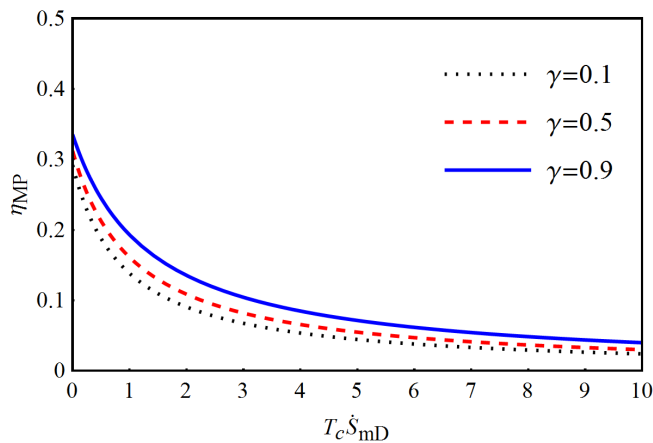}
	\caption{The efficiency at maximum power (obtained from Eq. (\ref{2al})) is plotted as function of the minimum dissipation for different values of $\gamma=0.1$ for dotted curve, $\gamma=0.5$ for dashed curve, $\gamma=0.9$ for solid curve. Here, we set $\eta_C=0.5$, $\eta_{ME}=0.45$, $P_{MP}=0.5$ and $P_{ME}=0.1$.}
	\label{fig:dissi}
\end{figure}

To analyze the dissipation effects on the efficiency at maximum power, we plotted the $\eta_{MP}$ obtained from Eq. (\ref{2al}) as function of the minimum dissipation $T_c\dot{S}_{mD}$ for different values of $\gamma$ and for given values of $P_{MP}$, $P_{ME}$ and $\eta_{ME}$ in Fig. \ref{fig:dissi}. The efficiency at maximum power decreases as the minimum dissipation increases. However, the $\eta_{MP}$ increases with increase in $\gamma$.

\begin{figure}[hbtp]
	\includegraphics[width=8.7cm]{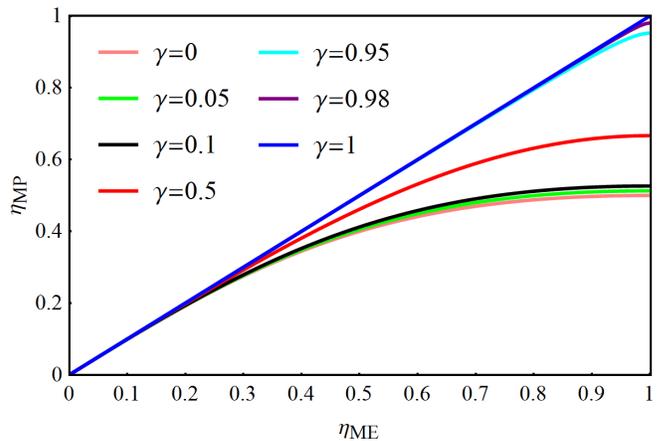}
	\caption{The efficiency at maximum power Eq. (\ref{2aq}) is plotted as function of the maximum efficiency for different values of $\gamma$. Here, the pink curve for $\gamma=0.01$, green curve for $\gamma=0.05$, black curve for $\gamma=0.1$, red curve for $\gamma=0.5$, cyan curve for $\gamma=0.9$, purple curve for $\gamma=0.95$ and blue curve for $\gamma=0.98$. Here, we set $\eta_C=1$}
	\label{fig:tretame}
\end{figure}

For time-reversal symmetric or anti-symmetric case, we plotted the $\eta_{MP}$ (Eq. (\ref{2aq})) as function of $\eta_{ME}$ for different values of $\gamma$ is shown in Fig. \ref{fig:tretame}. The efficiency at maximum power increases with the maximum efficiency as observed earlier in the general case. Interestingly, for a given value of  $\eta_C$, we observe that the linear behavior of $\eta_{MP}$ with $\eta_{ME}$ for higher values of $\gamma$ in contrast to the general case as shown in Fig. \ref{fig:etame}. However, at the asymmetrical dissipation limits with the reversible conditions ($\eta_{ME}=\eta_C$ and $P_{ME}=0$), the $\eta_{MP}$ in Eq. (\ref{2ah}) and the $\eta_{MP}$ in Eq. (\ref{2aq}) have the same lower and upper bounds. 

\begin{figure}[hbtp]
	\includegraphics[width=8.8cm]{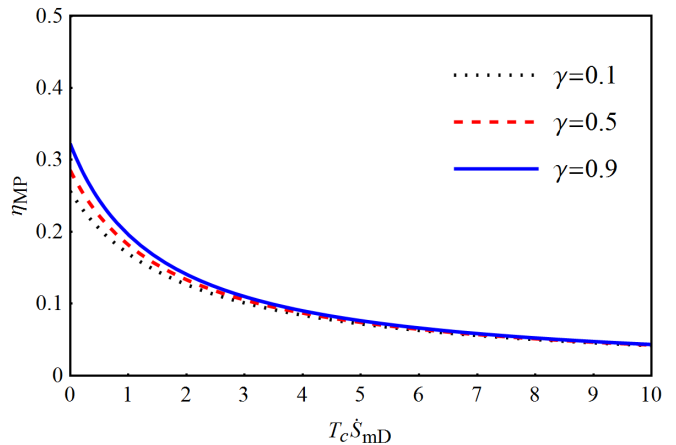}
	\caption{The efficiency at maximum power from Eq. (\ref{2df}) is plotted as function of the dissipation for different values of $\gamma$, where $\gamma=0.1$ for dotted curve, $\gamma=0.4$ for dashed curve, $\gamma=0.7$ for solid curve. Here, we set $\eta_C=1$ and $P_{MP}=1$.}
	\label{fig:trdissi}
\end{figure}

For time-reversal symmetric or anti-symmetric case, we plotted the efficiency at maximum power obtained from Eq. (\ref{2df}) as function of the minimum dissipation for various values of $\gamma$ and fixed $P_{MP}$ in Fig. \ref{fig:trdissi}. We did not see any appreciable changes in the behavior of decrease in $\eta_{MP}$ with the minimum dissipation as compared to the general case (see Fig. \ref{fig:dissi}).

Although the general relations derived in recent years Ref. \cite{jian042,baue042,proe220} along with our results need to be verified by experiments, analyzing the fundamental benchmark quantities $\eta_{ME}$ $\eta_{MP}$, $P_{ME}$ and  $P_{MP}$ is important for heat engines operating in finite-time cycle. It is also very important to understand the inter-connections among these fundamental performance parameters of the heat engines for theoretical as well as practical aspects. All these results may guide us to design a real heat engine with optimal outcome.

\section{Conclusion}

We studied the minimally nonlinear irreversible heat engine in the generalized framework with and without invoking the symmetry or anti-symmetry of the Onsager coefficients and obtained the general relations between the maximum power, maximum efficiency and the  minimum dissipation. For the asymmetric dissipation limits, we get the lower and upper bounds for those relations in which the lower bound is the same as the relations obtained recently for the linear irreversible heat engines \cite{proe220}. 
	
When the minimum dissipation $\dot{S}_{mD}=0$, we get the efficiency at maximum power as $\eta_{MP}^{mD}=\eta_C/(2-\gamma \eta_C)$, which was obtained previously for a thermoelectric heat engine and stochastic heat engine \cite{aper041,schm200}. Additionally, in the asymmetrical dissipation limits, we get the lower and upper bounds as $\eta_C/2\leq\eta_{MP}^{mD}\leq\eta_C/(2-\eta_C)$, which were obtained earlier for the low-dissipation Carnot heat engine and the minimally nonlinear irreversible heat engine with tight-coupling condition at the asymmetrical dissipation limits \cite{espo150,izum100}. 
Since our results can be applied to a three-terminal thermoelectric heat device in the nonlinear regime with broken time-reversal symmetry \cite{aper041,bran070}, which we will consider it as part of our future work.

It will be interesting to derive such a general relations for the other models of heat engines, such as the Feynman ratchet, information engine, quantum heat engines. Unlike the power output of the heat engine, the optimization parameter for refrigerator is not yet identified in general \cite{izum052,de010,izum10005,hu062,aper400}. However, the above general relations can also be obtained, if one can identify the suitable optimization criteria for refrigerator equivalent to the power output of the heat engine.

\begin{center}
\textbf{ACKNOWLEDGEMENT}
\end{center}
One of the author I.I would like to thank K. Proesmans for valuable discussions and also thank K. R. S. Preethi Meher, N. Barani Balan and H. Saveetha for helpful suggestions. We thank the anonymous referees for their constructive comments and criticism.

\end{document}